\documentclass[a4paper]{panl}
\usepackage{cite}
\usepackage{wrapfig}
\usepackage{graphicx}
\usepackage{amssymb}
\usepackage{amsfonts}
\usepackage{amsmath}
\usepackage{longtable}
\usepackage{rotating}
\usepackage{lscape}
\usepackage{epsfig}
\usepackage{multirow}

\usepackage{slashed}
\usepackage{xcolor}
\usepackage{bm}

\newcommand{\nt}[1]{\lefteqn{#1}\slash}

\usepackage{lineno}

\originalTeX
\begin{document}

\title{Renormalon-chain contributions to two-point correlators of nonlocal quark currents
}
\maketitle
\authors{S.\,V.\,Mikhailov$^{a,}$\footnote{E-mail: mikhs@theor.jinr.ru},
N.\,Volchanskiy$^{a,b,}$\footnote{E-mail: nikolay.volchanskiy@gmail.com}}
\setcounter{footnote}{0}
\from{$^{a}$\,Bogoliubov Laboratory of Theoretical Physics, JINR, 141980 Dubna, Russia}
\from{$^{b}$\,Research Institute of Physics, Southern Federal University,\\
                Prospekt Stachki 194, 344090, Rostov-na-Donu, Russia}


\begin{abstract}
%
We calculate, within massless QCD, a two-point correlator of nonlocal (composite) vector quark currents with arbitrary-length chains of the simplest fermion loops being inserted into gluon lines. Within the large $n_f$ (or large $\beta_0$) approximation, the correlator defines a perturbative contribution to the leading-twist distribution amplitudes for light mesons. Our results are consistent with a number of special cases in the literature. We consider functionals of the correlator, which are important for the phenomenology, and their properties as function series.
\end{abstract}
\vspace*{6pt}

\noindent
PACS: 11.15.Pg; 11.25.Db; 12.38.-t; 12.38.Bx 

\setcounter{footnote}{0}

\label{sec:intro}
\section*{Introduction}

We consider two-point correlators $\Pi_{n}(x,y;L)$ of nonlocal vector quark currents within large-$\beta_0$ approximation to massless perturbative QCD\footnote{We work in QCD with $n_f=3$ massless quark flavors; $N_c=3$ is the number of colors; the Casimir invariants are $C_A=3$ and $C_F=4/3$; $\beta_0 = \frac{11}{3} C_A - \frac43 T_F n_f = 9$ is the one-loop $\beta$ function coefficient; $T_F = \frac12$; $a_s = \alpha_s /(4\pi)$ is the coupling constant.} in $\overline{\text{MS}}$ scheme,
\begin{multline}\label{eq:cor-def}
-i\frac{a_s}{\pi^2} N_c C_F A^n \Pi_{n}(x,y;L) = \int \mathrm{d}^D\eta \, e^{ip\eta} \langle 0|\hat{\mathrm{T}}\left[J^\dagger(\eta;x)J(0;y) \right]|0 \rangle
\\{}
 = \vcenter{\hbox{\includegraphics[width=.18\textwidth]{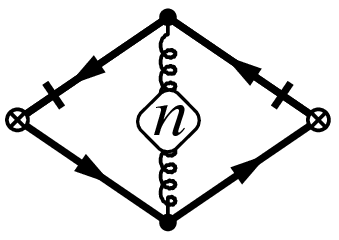}}}
 + \vcenter{\hbox{\includegraphics[width=.18\textwidth]{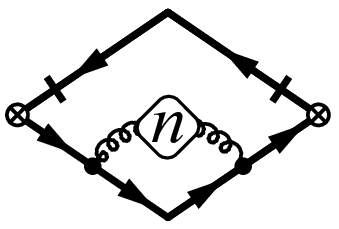}}}
 + \vcenter{\hbox{\includegraphics[width=.18\textwidth]{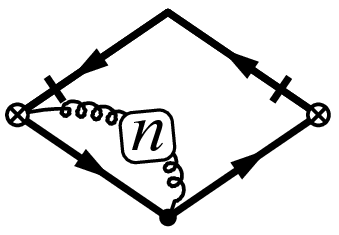}}}
 + \vcenter{\hbox{\includegraphics[width=.18\textwidth]{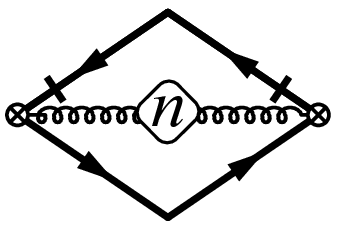}}}
 + \dots
\end{multline}
Here, $L=\ln(-p^2/\mu^2)$ with $p$ being an external momentum and $\mu$ being the renormalization scale and the constant $ A = \frac43 a_s T_F n_f$ can be replaced by $-a_s \beta_0$ as prescribed by the naive nonabelization trick.
\setcounter{footnote}{1}
In Eq.~\eqref{eq:cor-def}, the nonlocal vector quark current $J(\eta;x)$ is defined as the inverse Mellin transform $\hat{\mathtt{M}}^{-1}$ of a quark bilinear involving  the $N$th derivative of a quark field operator\footnote{Note that, in this paper, arguments of the Mellin transform are underlined, i.e.\ $f(\underline{a}) = \hat{\mathtt{M}} f(x) = \int_0^1 \mathrm{d}x \, f(x) x^a $.}:
\begin{align}\label{eq:currents-def1}
	J(\eta;x) = \hat{\mathtt{M}}^{-1}J(\eta;\underline{N}),
	\qquad
	J(\eta;\underline{N}) = \bar{d}(\eta) \nt{\tilde{n}} \left(i\tilde{n}\nabla\right)^N u(\eta),
\end{align}
where $x$ is a Bjorken fraction, $\eta$ is a space-time point, $\nabla_\mu=\partial_\mu -i g t_a A^a_\mu$ is the QCD covariant derivative, $\tilde{n}^\mu$ is a light-like vector, $\tilde{n}^2=0$. In QCD, the nonlocal current \eqref{eq:currents-def1} emerges naturally in the description of hard exclusive processes---its projection on a helicity-zero meson state gives twist-2 distribution amplitude (DA) of a meson. DA accumulates information about long-distance dynamics of partons constituting the meson and carrying a fraction $xp$ of the meson momentum $p$.

Within the approach of QCD sum rules (SR), the Borel transform $\hat{\mathtt{B}}$ of the correlator~\eqref{eq:cor-def} determines the perturbative contributions into meson DA, 
\begin{gather}\label{eq:DA-def}
\text{DA}(x;L_\text{B}) = \hat{\mathtt{B}} \frac{a_s}{\pi^2} N_c C_F \sum_{n \geqslant 0} A^n \Pi_{n}(x,\underline{0};L),
\quad
\Pi_n(x,\underline{0};L)=\int_0^1\Pi_n(x,y;L)\,\mathrm{d}y,
\end{gather}
where $L_\text{B} = \ln(M_\text{B}^2/\mu^2)$ is the logarithm of the Borel parameter $M_\text{B}$. In the approximation of large $\beta_0$ (or $n_f$), the pQCD part of SR is completely determined by diagrams \eqref{eq:cor-def} of two-loop topology with gluon lines dressed by one-loop fermion insertions---renormalon chains
\begin{gather*}
\vcenter{\hbox{\includegraphics[width=.2\textwidth]{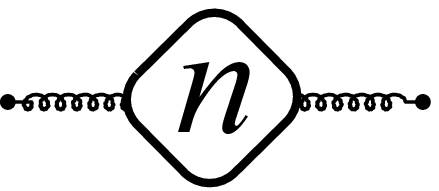}}}
 = \underbrace{\vcenter{\hbox{\includegraphics[width=.7\textwidth]{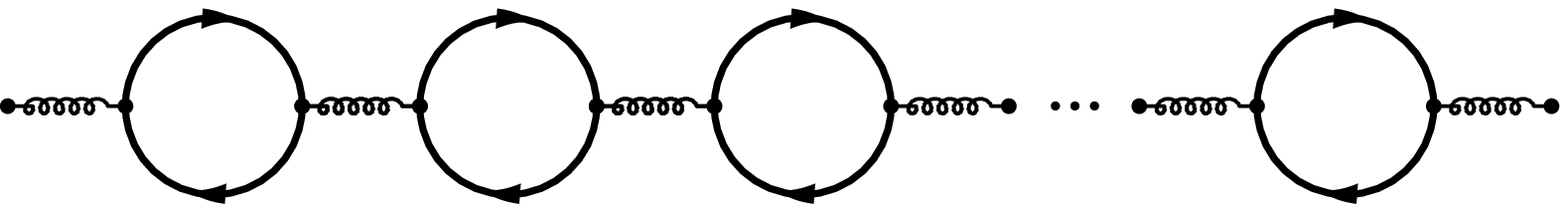}}}}_{n}.
\end{gather*}


\label{sec:x0-correlator}
\section{The generating function for the correlator $\Pi_n(x,\underline{0};L)$}

Let us now discuss the properties of $\Pi_n(x,\underline{0};L)$, which is the two-point correlator of one nonlocal and one local quark current. The general expression for the corresponding diagram of two-loop topology \eqref{eq:cor-def} with nonlocal vertices and arbitrary exponent of internal line propagator was derived in \cite{Mikhailov:2018udp}. This ``kite'' diagram can be represented in terms of the hypergeometric functions $_3F_2(x)$ and  $_3F_2(\bar{x})$, $\bar{x} =1-x$. Due to this, the sequence of $\Pi_n(x,\underline{0};L)$ can be condensed as two generating functions, an exponential $\Pi_{n}'$ and an ordinary $\Pi_{n}''$:
\begin{align}\label{eq:pix0}
\Pi_{n}(x,\underline{0};L) = \Pi_{n}'(x,\underline{0};L) + \Pi_{n}''(x,\underline{0};L),
\end{align}
\begin{multline}\label{eq:pix0'}
\sum_{n \geqslant 0} \frac{A^n}{n!} \frac{\mathrm{d}}{\mathrm{d}L} \Pi_{n}'(x,\underline{0};L)\\
{}= \mathop{\hat{\mathbf{S}}} \Biggl\{ \frac{e^{A(L-5/3)} x^A}{A^2(1+A)(2+A)} \Biggl[ -\bar{x} ( A + 4 x ) + 2 x \bar{x} \frac{(\pi A)^2 \cot(\pi A)}{x^A\sin(\pi A)}
\\
 {}+ A x (2\bar{x}+A) \text{B}_{\bar{x}}(A,1-A) + \frac{2x^2\bar{x}A^2}{(1+A)^2} {}_3F_2  \left(  \left.\begin{matrix}
																						1,\, 1,\, 1+A \, \\
																						2+A,\, 2+A \,
																	\end{matrix} \right\rvert x \right)
 \Biggr] \Biggr\},
\end{multline}
\begin{align}\label{eq:pix0''}
&\sum_{n \geqslant 0} A^n \frac{\mathrm{d}}{\mathrm{d}L} \Pi_{n}''(x,\underline{0};L) = - \frac1{A} \int_0^A \mathrm{d}a \, F(x;a),
\end{align}
where
\begin{gather}
F(x;a) =  \frac{1}{2a}\int_0^1 \mathrm{d}y\, y\bar{y} \left[ \frac{ V(x,y;a)}{ h_1(a) } - V(x,y;0) \right]_{+(x)},
\\
h_1(a) = \frac{(1-a) \Gamma(1+a) \Gamma^3(1-a)}{(1-2a/3) (1-2a) \Gamma(1-2a)},
\\
V(x,y;a) = 2 \mathop{\hat{\mathbf{S}}} \left[ \Theta(y>x) \left( \frac{x}{y} \right)^{1-a} \left( 1-a + \frac{1}{y-x} \right) \right].
\end{gather}
Here, the function $h_1(\varepsilon)$ comes from $\varepsilon$-dependence of the simplest quark loop in the gluon propagator ($D=4-2\varepsilon$ is the space-time dimension), $V(x,y;a)$ is a generalization of one-loop ERBL evolution-equation kernels, $f(x,y)_{+(x)} = f(x,y) - \delta(x-y) f(\underline{0},y)$ is the plus distribution, and $\mathop{\hat{\mathbf{S}}} \left[ f(x,y) \right] = f(x,y) + f(\bar x, \bar y)$. Note that in the scope of this paper we are not interested in the nonlogarithmic term of the correlator, $\Pi_{n}(x,\underline{0};L=0)$, since $\hat{\mathtt{B}}\left(  \text{const} \right)= 0$. The part of the correlator that is represented as the ordinary generating function \eqref{eq:pix0''} is related to the counterterms in the nonlocal vertex.

From \eqref{eq:pix0}--\eqref{eq:pix0''}, we can derive explicit coefficients of the $L$-expansion of the correlator
\begin{gather}
\Pi_n(x,\underline{0};L) = (-1)^n n! \sum_{k=0}^{n+1} \frac{(-L)^k}{k!} \Pi_n^k(x,\underline{0}).
\end{gather}
The highest degree term $\Pi_n^{n+2}(x,\underline{0})$ is equal to 0 because of the gauge symmetry and current conservation. The first nonvanishing coefficient reads
\begin{align}
\Pi_{n}^{n+1}(x,\underline{0}) &{}= \frac12 \mathop{\hat{\mathbf{S}}}  \left\{ x \ln x + \delta_{0,n} \left[ - x \ln x + \frac12 x \bar{x} \left( \frac{\pi^2}{3} - 5 - \ln^2 \frac{x}{\bar{x}} \right) \right] \right\},
\end{align}
which is in agreement with the previous calculations. The following terms grow increasingly lengthy to be written out in proceedings. Nevertheless, what (highest transcendency) types of functions they are expressed in terms of can still be specified:
\begin{gather*}
\Pi_{n > 0}^n(x,\underline{0}) \sim \mathop{\hat{\mathbf{S}}} \mathop{\mathrm{Li}_{3}} x + \text{simpler polylogarithms},
\quad
\Pi_{n > 1}^{n-1}(x,\underline{0}) \sim \mathop{\hat{\mathbf{S}}} \mathop{\mathrm{Li}_{4}} x + \dots,
\\
\Pi_n^{k>0}(x,\underline{0}) \sim \mathop{\hat{\mathbf{S}}} \mathrm{H}_\text{\bm{${\mu}$}} (x) + \dots,
 \qquad
  \text{\bm{${\mu}$}}=\mu_1,\dots \mu_r: \;\mu_i>0,\; \sum \mu_i = n-k+3,
\end{gather*}
where $\mathrm{H}_\text{\bm{${\mu}$}} (x)$ are harmonic polylogarithms \cite{Remiddi:1999ew},
\begin{gather*}
{\mathop{\mathrm{H}}}_\text{\bm{${\mu}$}}(z) = \sum_{\sigma} z^{m_1} \prod_{i=1}^r \frac{1}{m_i^{\mu_i}},
\qquad \lvert z \rvert < 1,
\\
\text{\bm{${\mu}$}}=\mu_1,\dots \mu_r,
\qquad
\sigma = \left\{ \forall m_i \in \mathbb{N}, i=1,\dots r: m_1 > m_2 > \dots > m_r > 0 \right\}.
\end{gather*}

\begin{figure}[t]
\begin{center}
\includegraphics[width=0.325\textwidth]{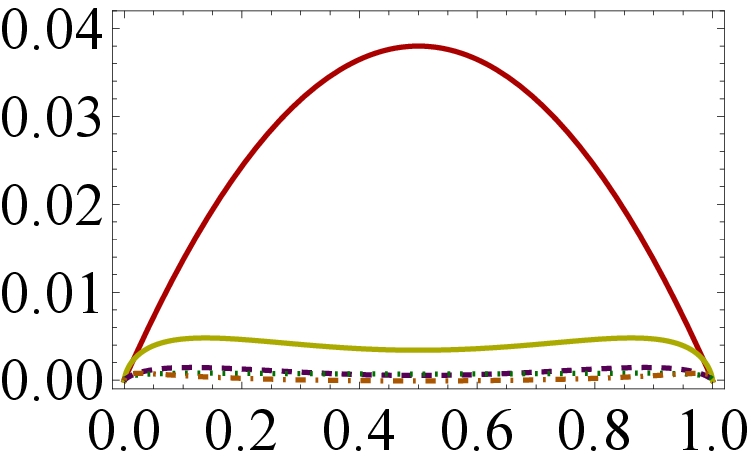}
\includegraphics[width=0.325\textwidth]{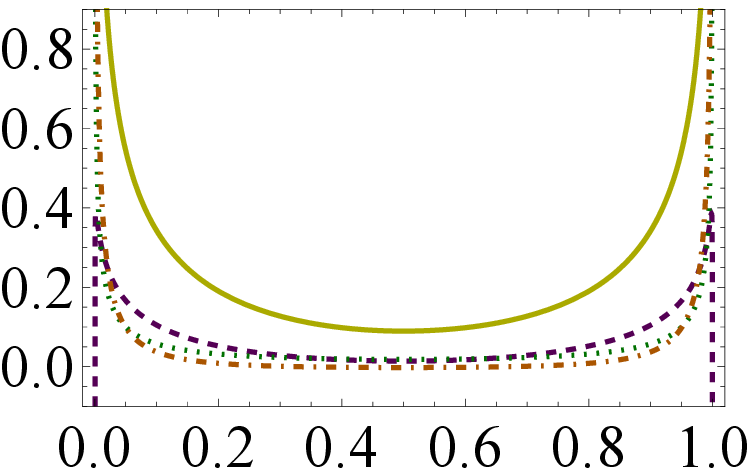}
\vspace{-3mm}
\caption{Left panel: LO (\textcolor[cmyk]{0,1,.91,.37}{\bm{$-$}}), NLO (\textcolor[cmyk]{0.19,0.16,.82,.18}{\bm{$-$}}), $\beta_0$N$^2$LO (\textcolor[cmyk]{0.52,0.83,.2,.43}{\mbox{\textbf{-\,-}}}), $\beta_0^2$N$^3$LO (\textcolor[cmyk]{1,0,.98,.6}{\mbox{$\bm{\cdot\cdot\cdot}$}}), and $\beta_0^3$N$^4$LO (\mbox{\textcolor[cmyk]{0,0.62,1,.4}{$\bm{\text{\textbf{-}}\cdot\text{\textbf{-}}}$}}) contributions to DAs for pseudoscalar or longitudinally polarized vector mesons. 
Right panel: the ratios NLO/LO (\textcolor[cmyk]{0.19,0.16,.82,.18}{\bm{$-$}}), $\beta_0$N$^2$LO/LO (\mbox{\textcolor[cmyk]{0.52,0.83,.2,.43}{\textbf{-\,-}}}), and $\beta_0^2$N$^3$LO/LO (\mbox{\textcolor[cmyk]{1,0,.98,.6}{$\bm{\cdot \cdot \cdot}$}}), and $\beta_0^3$N$^4$LO/LO (\mbox{\textcolor[cmyk]{0,0.62,1,.4}{$\bm{\text{\textbf{-}}\cdot\text{\textbf{-}}}$}}). All curves are for the case of $L_\text{B}=0$, $\alpha_s (\mu^2 = 1 \text{ GeV}^2) \approx 0.49$. }
\end{center}
\labelf{fig:x0}
\vspace{-5mm}
\end{figure}

Fig.~\ref{fig:x0} shows several lowest-order contributions to meson DAs obtained from Eqs.~\eqref{eq:pix0}--\eqref{eq:pix0''} with the help of \eqref{eq:DA-def} and the Borel transform
\begin{gather}\label{eq:borel}
\hat{\mathtt{B}} \left[ f(t) \right](\mu) = \lim_{\begin{subarray}{c} t=n\mu \\ n\to \infty\end{subarray}} \frac{(-t)^n}{\Gamma(n)} \frac{\mathrm d^n }{\mathrm d t^n}f(t),
\qquad
\hat{\mathtt{B}} e^{AL} = - \frac{A e^{A L_\text{B}}}{\Gamma(1-A)}.
\end{gather}
These curves exhibit different behavior for the intermediate values of the Bjorken variable $x$, where they decrease sequentially from LO to N$^4$LO, and at the endpoints, where their ratios become singular. The vicinity of endpoints is quantitatively important for DAs of pseudoscalar and longitudinally polarized vector mesons. Therefore, it makes sense to look at two integral characteristics of the correlators---their zeroth and inverse moments, $\Pi_{n}(\underline{0},\underline{0};L)$ and $\Pi_{n}(\underline{-1},\underline{0};L)$. They are formed mostly by the intermediate and near-endpoint values of the $x$-dependent correlator, respectively.


\subsection{The zeroth moment $\Pi_{n}(\underline{0},\underline{0};L)$.~}

The derivative of the zeroth moment with respect to $L$ is proportional to the Adler function of QCD. The corresponding exponential generating function reads
\begin{multline}\label{eq:Pi(0,0)}
\sum_{n \geqslant 0} \frac{A^n}{n!} \frac{\mathrm{d}}{\mathrm{d}L} \Pi_{n}(\underline{0},\underline{0};L)
 = \frac{e^{A(L-5/3)}}{6(1+A)(2+A)} \Biggl[ - \psi_1\left(\frac{4+A}{2}\right) + \psi_1\left(\frac{3+A}{2}\right)
 \\{} + \psi_1\left(\frac{2-A}{2}\right) - \psi_1\left(\frac{1-A}{2}\right) \Biggr],
\end{multline}
where $\psi_1$ is the trigamma function. The expression \eqref{eq:Pi(0,0)} agrees with the calculation \cite{citeulike:10762103} of the correlator and its anomalous dimension for $n=0,\, 1, \,2, \, 3$. Also, it coincides with the Adler function $D(a_s,L)$ from \cite{Ball:1995ni} for $n=2, \, 3$ and all-order $D(a_s,L)$ from \cite{1993ZPhyC..58..339B,Broadhurst:1993ru}. 

The behavior of the Borel transform of $\Pi_{n}(\underline{0},\underline{0};L)$ is depicted in Fig.~\ref{fig:moms}. This asymptotic series should be truncated at $n=3$ where it becomes divergent and bursts into factorial growth at $n>10$.


\subsection{The inverse moment $\Pi_{n}(\underline{-1},\underline{0};L)$.~}

The two generating functions for the inverse moment can be written as
\begin{gather}
 \Pi_{n}(\underline{-1},\underline{0};L) = \Pi'_n(\underline{-1},\underline{0};L)+\Pi''_n(\underline{-1},\underline{0};L),
\notag \\
\sum_{n \geqslant 0} \frac{A^n}{n!} \frac{\mathrm{d}}{\mathrm{d}L} \Pi'_{n}(\underline{-1},\underline{0};L)
 = \frac{e^{A(L-5/3)}}{2(1+A)(2+A)} \Biggl[ \psi_1\left(\frac{2-A}{2}\right) - \psi_1\left(\frac{1-A}{2}\right) \Biggr],
\notag\\
\sum_{n \geqslant 0} A^n \frac{\mathrm{d}}{\mathrm{d}L} \Pi''_{n}(\underline{-1},\underline{0};L)= -\frac{1}{A} \int_0^A \mathrm{d}a F(\underline{-1},a),
\end{gather}
where
\begin{align*}
F(\underline{-1},a) = \frac{\Gamma(4-2a)}{6\Gamma(2-a)^2 \Gamma(3+a)} \left\{ \frac{5+6a-5a^2}{\Gamma(3-a)} + \frac{(1+2a) [\gamma_\text{E} + \psi(1-a) ]}{a \Gamma(1-a)}\right\}.
\end{align*}
Fig.~\ref{fig:moms} illustrates the behavior of the sequence of borelized $\Pi_{n}(\underline{-1},\underline{0};L)$ that can be obtained with the help of \eqref{eq:borel}. The series becomes factorially divergent at $n=4$.
\begin{figure}[t]
\begin{center}
\includegraphics[width=0.48\textwidth]{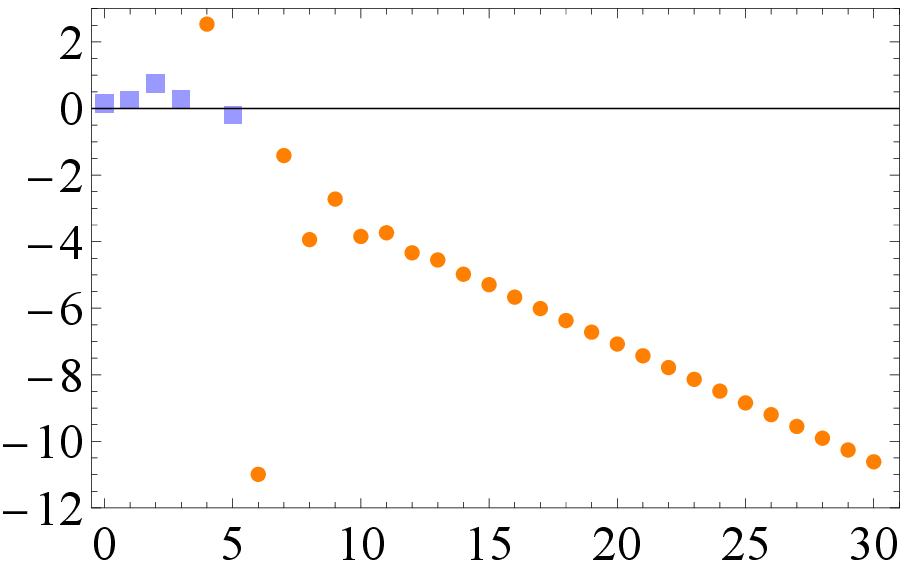}
\includegraphics[width=0.48\textwidth]{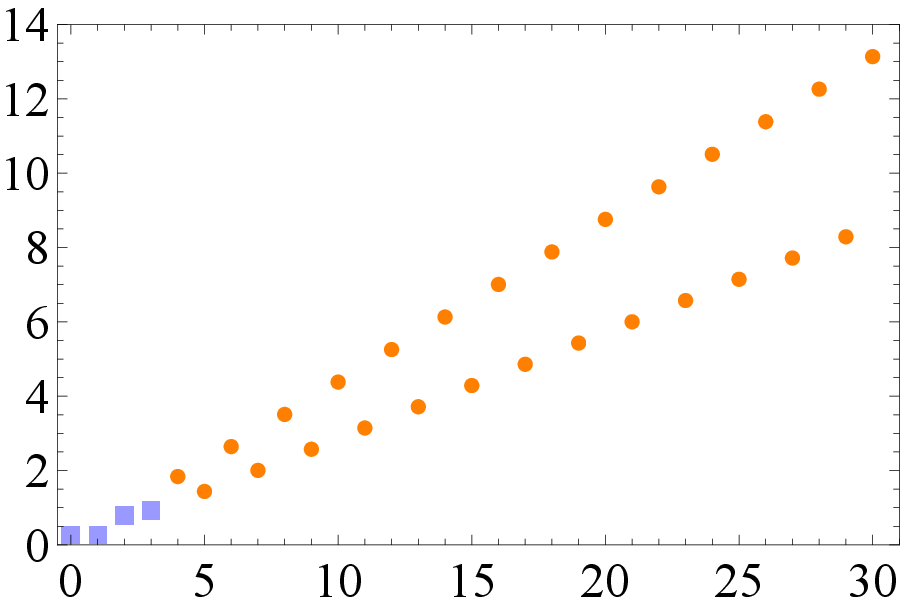}
\vspace{-3mm}
\caption{The ratio $R_n(\underline{N}) = -a_s \beta_0 \hat{\mathtt{B}} \Pi_n(\underline{N},\underline{0};L) / \hat{\mathtt{B}} \Pi_{n-1}(\underline{N},\underline{0};L)$ for $N=0$ (left panel) and $N=-1$ (right panel); $R_0$ is defined as the ratio of 2-loop and 1-loop correlators, $R_0(\underline{0}) = 3 a_s C_F$ and $R_0(\underline{-1}) = 5 a_s C_F$ \cite{Mikhailov:1988nz, Mikhailov:2020tta}. Blue squares are for $R_n \leqslant 1$. All free parameters are the same as in Fig.~\ref{fig:x0}}
\end{center}
\labelf{fig:moms}
\vspace{-5mm}
\end{figure}

\section{Conclusion}

We have evaluated the correlator of vector nonlocal quark currents of order $a_s^{n+1}\beta_0^n$ in QCD, $n \geqslant 0$. The lower Mellin moments of the correlator have been calculated. The zeroth moment as well as some other fixed-order special cases agree with previous calculations in the literature. Generating functions for the correlator and its moments have been constructed. The correlator at any fixed order $a_s^{n+1}\beta_0^n$ can be expressed in terms of harmonic polylogarithms of weight not higher than $n+2$. We briefly discussed how the higher order radiative corrections affect DA behavior.

\textbf{Acknowledgements} NV was supported by the Russian Science Foundation grant No-18-12-00213-P.


\end{document}